\documentclass[sigconf]{acmart} 
\AtBeginDocument{%
  }

\setcopyright{acmlicensed}

\copyrightyear{2026}
\acmYear{2026}
\setcopyright{cc}
\setcctype{by}
\acmConference[SIGIR '26]{Proceedings of the 49th International ACM SIGIR Conference on Research and Development in Information Retrieval}{July 20--24, 2026}{Melbourne, VIC, Australia}
\acmBooktitle{Proceedings of the 49th International ACM SIGIR Conference on Research and Development in Information Retrieval (SIGIR '26), July 20--24, 2026, Melbourne, VIC, Australia}
\acmDOI{10.1145/3805712.3808474}
\acmISBN{979-8-4007-2599-9/2026/07}

\begin{document}

\title{Scaling and Stabilizing Large-Scale Embedding-Based Retrieval}

\settopmatter{authorsperrow=4}

\author{Zhen Yang}
\authornote{Work done while at Walmart.}
\authornote{Both authors contributed equally to this research.}
\affiliation{%
  \institution{Walmart Global Tech}
  \city{Sunnyvale}
  \country{USA}
}
\email{njuyangz@gmail.com}

\author{Juexin Lin}
\authornotemark[1]
\authornotemark[2]
\affiliation{%
  \institution{Walmart Global Tech}
  \city{Sunnyvale}
  \country{USA}
}
\email{linjuexin@gmail.com}

\author{Hongwei Shang}
\affiliation{%
  \institution{Walmart Global Tech}
  \city{Sunnyvale}
  \country{USA}
}
\email{hongwei.shang@walmart.com}

\author{Kaihao Li}
\affiliation{%
  \institution{Walmart Global Tech}
  \city{Sunnyvale}
  \country{USA}
}
\email{kaihao.li@walmart.com}

\author{Feng Liu}
\authornotemark[1]
\affiliation{%
  \institution{Walmart Global Tech}
  \city{Sunnyvale}
  \country{USA}
}
\email{liu.f@pku.edu.cn}

\author{Satya Chembolu}
\affiliation{%
  \institution{Walmart Global Tech}
  \city{Sunnyvale}
  \country{USA}
}
\email{satya.chembolu@walmart.com}

\author{Xunfan Cai}
\affiliation{%
  \institution{Walmart Global Tech}
  \city{Sunnyvale}
  \country{USA}
}
\email{xunfan.cai@walmart.com}

\author{Xinyi Liu}
\affiliation{%
  \institution{Walmart Global Tech}
  \city{Sunnyvale}
  \country{USA}
}
\email{xinyi.liu@walmart.com}

\author{Cun	Mu}
\affiliation{%
  \institution{Walmart Global Tech}
  \city{Sunnyvale}
  \country{USA}
}
\email{cun.mu@walmart.com}

\author{Tony Lee}
\authornotemark[1]
\affiliation{%
  \institution{Walmart Global Tech}
  \city{Sunnyvale}
  \country{USA}
}
\email{tonyleexyz@gmail.com}

\author{Ciya Liao}
\authornotemark[1]
\affiliation{%
  \institution{Walmart Global Tech}
  \city{Sunnyvale}
  \country{USA}
}
\email{liao_ciya@yahoo.com}

\renewcommand{\shortauthors}{Zhen Yang et al.}

\begin{abstract}
Embedding-based retrieval (EBR) is foundational to large-scale e-commerce search, yet its effectiveness is often constrained by the quality of training signals and the representational capacity of the encoder. Standard dual-encoders suffer from a training-inference gap: they are optimized on narrow candidate pools but must discriminate against hundreds of millions of items during inference. Furthermore, while transitioning to higher-capacity backbones can mitigate this gap, simply replacing a mature model can lead to inconsistent retrieval behavior and a loss of the domain-specific knowledge established in previous iterations. In this paper, we present a unified pipeline deployed at Walmart that addresses both signal quality and model evolution. Our contributions are two-fold: (1) Hybrid Hard Negative Mining: We integrate Online Cross-Batch Sampling to increase negative diversity by an order of magnitude and Hybrid Offline Mining, which combines cross-encoder predictions with metadata heuristics to identify nuanced mismatches. (2) Legacy-Aware Distillation: We transition from DistilBERT to a higher-capacity GTE-base encoder. To ensure a smooth and superior transition, we introduce a Warm-Start Distillation technique that transfers domain-specific expertise from the legacy model to the new backbone. Validated through extensive offline experiments and online A/B testing, the proposed pipeline is deployed in live production, delivering a +7.34\% improvement in NDCG@5 and a +0.50\% lift in gross revenue.
\end{abstract}




\maketitle

\section{Introduction}

Embedding-based retrieval (EBR) using dual-encoder architectures and Approximate Nearest Neighbor (ANN)~\citep{johnson2019billion} search has become a foundational component of large-scale e-commerce platforms. By mapping queries and products into a shared dense vector space, EBR enables semantic matching that transcends traditional lexical constraints and addresses common issues like the vocabulary mismatch problem. Prior works ~\citep{magnani2022semantic,qiu2022pre,freymuth2025hierarchical} have demonstrated that EBR can significantly improve recall and user engagement over purely lexical approaches. As a result, EBR has been widely adopted in the search stacks of major platforms, including Walmart, Amazon, Taobao, and Facebook~\citep{magnani2022semantic,lin2024enhancing,rossi2024relevance,li2021embedding,muhamed2023web,he2023que2engage}. However, despite these semantic advantages of EBR, its effectiveness in production is often constrained by a training-inference gap and model capacity limits.


While in-batch sampling is efficient and widely adopted in e-commerce \citep{henderson2017efficient,magnani2022semantic,lin2024enhancing}, it is fundamentally constrained by the narrow diversity of a single batch. To expose models to harder distractors, offline mining methods have evolved from lexical heuristics \citep{karpukhin2020dense,robertson1994some} to ANCE-style periodic index refreshes \citep{xiong2020ance}, often augmented with domain-specific constraints \citep{lin2024enhancing,magnani2022semantic,meghwani2025hard}. However, because offline mining evaluates a massive global index to retrieve the most challenging candidates, it is highly susceptible to sampling false negatives. This noise degrades the quality of the training set. To scale negative diversity while preserving signal quality, we propose a dual-signal approach. Online, we implement cross-batch sampling \citep{qu2021rocketqa} to share representations across GPUs, exponentially expanding the negative pool to better approximate massive catalogs. Offline, we capture a broader, more diverse range of hard negatives by extending \citet{lin2024enhancing}'s framework, integrating a cross-encoder and metadata guardrails to safely mine semantically confusable negatives that traditional heuristics miss.

Enriched training signals expose a secondary bottleneck: model capacity. The representational capacity of models in industry like DistilBERT~\citep{sanh2019distilbert} often lags behind the state-of-the-art, which struggle to resolve the complex decision boundaries of harder negatives. Transitioning to a higher-capacity encoder like GTE-base \citep{li2023towards} mitigates this gap but introduces significant model-evolution risk. Even with standard fine-tuning on current data, a newly initialized high-capacity model often fails to outperform a smaller legacy model that has been iteratively refined over multiple generations. Starting "cold" from public checkpoints discards this accumulated domain knowledge, consistently leading to performance regression.
To safely execute this upgrade, we introduce a Legacy-Aware Warm-Start Distillation technique. While traditional knowledge distillation typically focuses on compressing a large teacher into a smaller student for inference efficiency \citep{hinton2015distilling,hofstatter2020improving}, recent literature has begun exploring "inverse" distillation to bootstrap larger architectures \citep{ruan2024idat}. Building on this emerging paradigm, we utilize the highly refined legacy model as a teacher for the larger, uninitialized backbone prior to full fine-tuning, we prevent performance instability.
With this warm start, our final model preserves historical domain expertise and ensures the higher-capacity model fully exploits the enriched signals.


To address these challenges, we propose a unified "Scaling and Stabilizing" framework deployed at Walmart:
\begin{itemize}
\item \textbf{Scaling Negatives through Hybrid Mining:} We implement Online Cross-Batch Sampling to expand the negative pool by an order of magnitude. This is augmented by Hybrid Offline Mining, which integrates a cross-encoder with metadata guardrails to identify nuanced mismatches.

\item \textbf{Stabilizing through Legacy-Aware Distillation:} We facilitate the transition to a higher-capacity encoder using a Warm-Start Distillation technique. By initializing the new student model to approximate the score distribution of the legacy model, we preserve domain-specific expertise while unlocking the capacity to exploit improved training signals.
\end{itemize}

The proposed system has been evaluated both offline and online, deployed to serve real user traffic, delivering consistent improvements across relevance metrics (NDCG@5 + 7.34\%) and key business metrics (gross revenue +0.50\%).

\section{Methodology}

\subsection{Model Architecture}

\subsubsection{Dense Retrieval Model}
\label{sec:dense}
We adopt the same dual-encoder model architecture and training objective as described in \citet{lin2024enhancing}. Queries and products (product title associated with attributes: product type, brand, color, age, gender) are independently encoded into dense vector representations using identical encoder architectures, and retrieval is performed via similarity search in the shared embedding space. To balance user interaction signals with semantic fidelity, we optimize a joint objective comprising an Engagement Loss ($\mathcal{L}_{Eng}$) and a Relevance Loss ($\mathcal{L}_{Rel}$). The formulation is defined as:
\begin{equation}
\mathcal{L}_{Eng_i} = - \sum_{j=1}^{N} \tilde{S}_{ij} \log \left( \frac{\exp(\cos(q_i, p_j) / \sigma)}{\sum_{k=1}^{N} \exp(\cos(q_i, p_k) / \sigma)} \right),
\end{equation}

\begin{equation}
\mathcal{L}_{Rel_i} = - \sum_{j=1}^{N} \tilde{R}_{ij} \log \left( \frac{\exp(\cos(q_i, p_j) / \tau)}{\sum_{k=1}^{N} \exp(\cos(q_i, p_k) / \tau)} \right),
\end{equation}

\begin{equation}
\mathcal{L}_{i} = \omega \cdot \mathcal{L}_{Eng_i} + (1 - \omega) \cdot \mathcal{L}_{Rel_i},
\end{equation}

\noindent where $N$ denotes the number of products, and $\cos(\cdot, \cdot)$ is the cosine similarity between two embeddings. $q_i$ and $p_j$ represent the embeddings for query $i$ and product $j$, respectively. $\tilde{S}'_{ij}$ and $\tilde{R}_{ij}$ are the normalized engagement and relevance labels. $\sigma$ and $\tau$ are learnable temperature parameters. Finally, $\omega$ is a hyperparameter empirically selected to balance the contribution of engagement and relevance. In our work, we set $\omega=0.5$.

\begin{table*}
\centering
\small
\caption{Examples of challenging offline hard negatives. In these instances, the heuristic Product Type (PT) successfully matches, yet the candidate remains strictly irrelevant due to critical attribute mismatches.}
\begin{tabular}{p{3.2cm} p{4.2cm} p{2.0cm} p{6.0cm}}
\toprule
\textbf{Query} & \textbf{Product} & \textbf{Class} & \textbf{Explanation} \\
\midrule
LG Television 32" 1080p  
& Samsung 65" 4K Smart LED TV  
& Irrelevant  
& PT matches, but major attribute mismatches: wrong brand, much larger size, and different resolution (4K vs. 1080p). \\

Revlon round hair brush  
& Conair Square Professional Hair Brush, Black
& Irrelevant  
& PT matches, but both the brand and brush shape/type differ from the query intent. \\
\bottomrule
\end{tabular}
\label{tab:qip_hard}
\end{table*}

\subsubsection{Relevance Cross-Encoder Model}
\label{sec:rel_model}
To assess query-product relevance, we employ a BERT-based \citep{devlin2019bert} cross-encoder model \cite{mehrdad2024large, puthenputhussery2025large, puthenputhussery2026large}. Similar to \citet{mehrdad2024large}, the model is formulated as a three-class classification task \citep{reddy2022shopping}, producing probabilities over irrelevant, semi-relevant, and exact match. To train this model, we first optimize a Mistral-7B teacher model \citep{Jiang2023Mistral7} on small-scale human-annotated data using cross-entropy loss. We then transfer this knowledge to the BERT student via distillation \citep{shang2025knowledge} using a Kullback-Leibler (KL) divergence objective \citep{kullback1951kullback}. For the model's input, the query is concatenated with the product's text attributes, including title, product type (PT), brand, color, and gender, with dedicated separator tokens. This combined sequence is fed into the BERT encoder and passed through non-linear layers with a softmax activation to obtain the final three-dimensional class probability distribution.

\subsection{Scaling Negatives through Hybrid Mining}

\subsubsection{Cross-Batch Online Hard Negatives}
While standard in-batch sampling is computationally efficient because embeddings can be reused~\citep{magnani2022semantic,lin2024enhancing}, it fundamentally restricts the negative pool to the local batch size $B$. To expose the model to a larger and more challenging distribution without incurring additional encoder forward passes, we implement online cross-batch negative sampling~\citep{wang2021cross}. 
Formally, let $q_i \in \mathbb{R}^d$ be the encoded representation of a query on a given device, and let $\mathbf{P}_{local} \in \mathbb{R}^{B \times d}$ denote the local item embedding matrix. In a standard setup, the contrastive loss denominator is strictly bounded by the local batch size $B$. Under the cross-batch framework, we utilize a differentiable all-gather communication primitive across $D$ distributed devices to construct a global item embedding matrix $\mathbf{P}_{global} \in \mathbb{R}^{(B \times D) \times d}$. This expands the negative candidate pool by allowing the similarity function to be computed against all parallel mini-batches, scaling the effective contrastive space from $B$ to $B \times D$. This substantially increases the diversity of negatives seen during training while preserving the memory footprint of standard in-batch execution.

 

\subsubsection{Offline Hard Negative Mining}

\begin{figure*}[t]
  \centering
  \includegraphics[width=\linewidth]{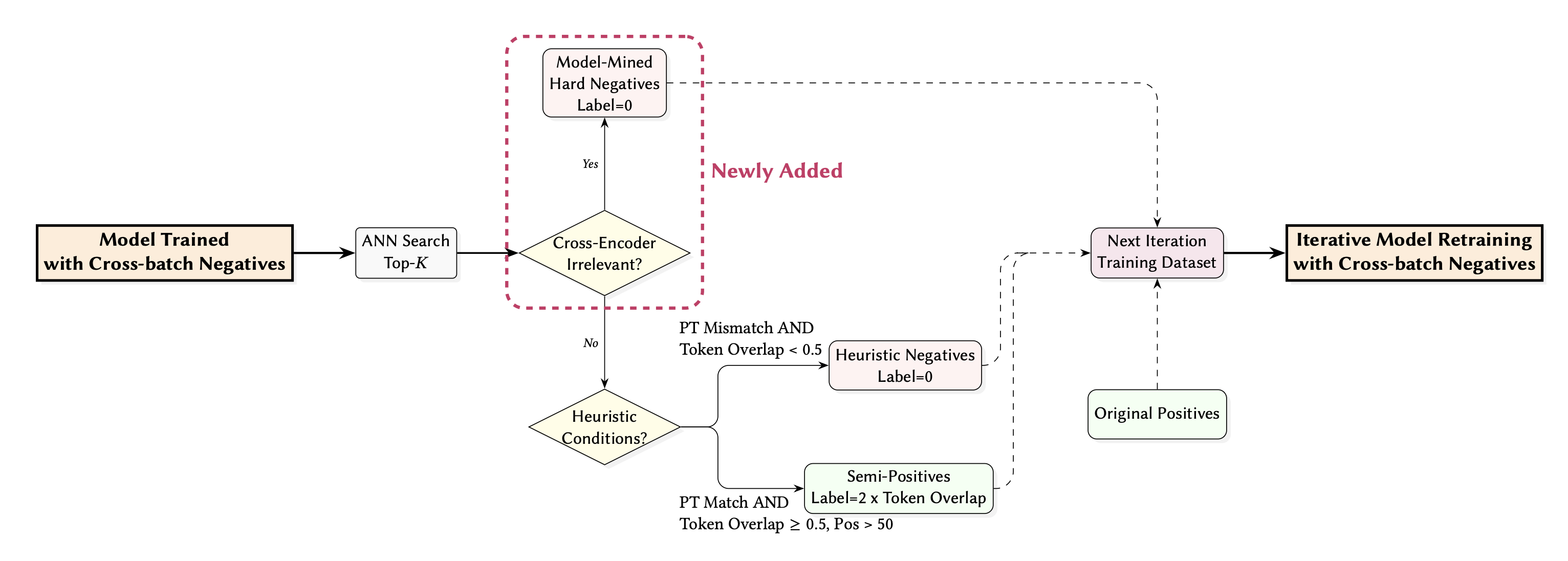}
\caption{Flowchart of the hybrid offline hard negative mining process.}
\label{fig:hard-negative-mining-ltr}
\end{figure*}

While heuristic offline mining based on Product Type (PT) mismatching successfully identifies basic negatives \citep{lin2024enhancing}, it fails to capture nuanced distractors—such as items that share the correct PT but mismatch on critical attributes like brand or size. As illustrated in Table~\ref{tab:qip_hard}, such cases are often excluded by PT-only filtering. To mine these challenging hard negatives and provide more informative training signals, we augment the iterative ANCE-style pipeline with the cross-encoder relevance model introduced in Section~\ref{sec:rel_model}. 
As detailed in Figure~\ref{fig:hard-negative-mining-ltr}, we first retrieve the top-$K$ candidates via an ANN search using the model from the previous training iteration. The cross-encoder then evaluates these candidates to isolate hard semantic distractors. Finally, the remaining items are classified into heuristic negatives or semi-positives based on PT matching and token overlap constraints, constructing a highly refined dataset for the next training cycle.

\subsection{Legacy-Aware Warm-Start Distillation}
\begin{figure}
  \centering
  \includegraphics[width=\linewidth]{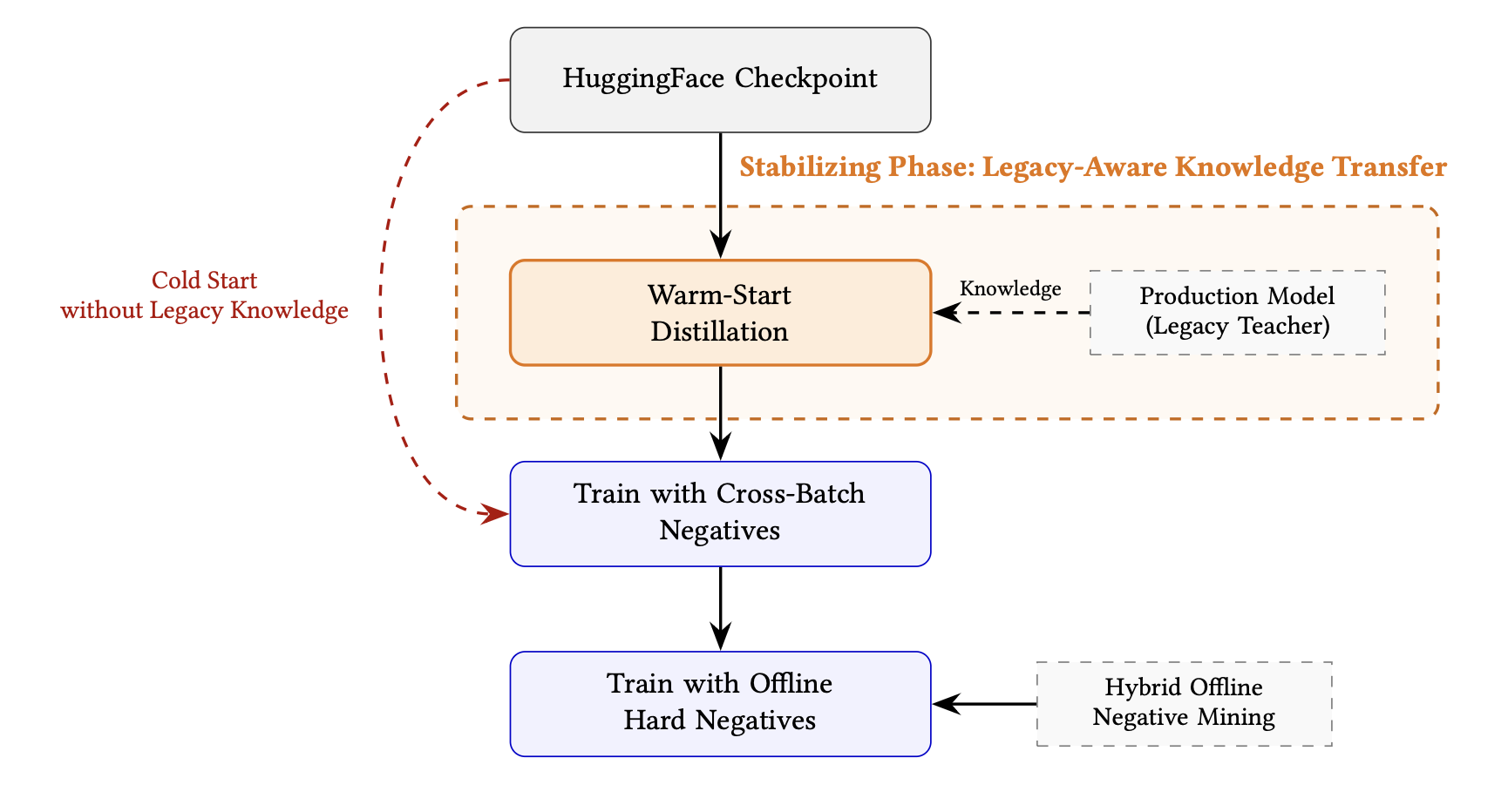}
\caption{Flowchart of legacy-aware warm-start distillation.}
\label{fig:legacy-aware-ws}
\end{figure}

Even when high-quality training signals are available, the representational capacity of the model often becomes the limiting factor. While lightweight legacy models like DistilBERT (teacher $M_{\mathcal{T}}$) struggle to resolve harder negatives, upgrading to a higher-capacity student $M_{\mathcal{S}}$ (e.g., GTE-base) introduces model-evolution risk. While standard fine-tuning on a recent snapshot of engagement data captures macro-level user intent, it does not fully encapsulate the platform's historical edge cases. The legacy model ($M_\mathcal{T}$) has undergone continual, iterative refinement over multiple generations—accumulating implicit knowledge of seasonal shifts, domain-specific vocabulary nuances, and historically mined hard negatives that are not entirely represented in a single, temporally bounded training snapshot. Therefore, initializing a higher-capacity model from scratch risks discarding this deeply embedded historical expertise.
To mitigate this domain shift and preserve the knowledge captured by the previous system, we introduce a Legacy-Aware Warm-Start distillation phase prior to full fine-tuning. As illustrated in Figure~\ref{fig:legacy-aware-ws}, this stabilizing phase bypasses the typical "cold start" from a public checkpoint by first transferring knowledge from the legacy production model. Following this Warm-Start Distillation, the newly initialized model proceeds to standard iterative training utilizing both the cross-batch and hybrid offline negatives.

We transfer accumulated historical expertise of $M_{\mathcal{T}}$ by minimizing the KL divergence between the temperature-scaled score distributions of the legacy production dual-encoder teacher and high-capacity student.  Here, $M_{\mathcal{T}}$ strictly refers to the original production baseline model (the DistilBERT architecture proposed by \citet{lin2024enhancing}), rather than the newly data refreshed checkpoint.
For a query $q$ and candidate set $P$, $$\mathcal{L}_{WS} = \sum_{p \in P} s_\mathcal{T}(q, p) \log \left( \frac{s_\mathcal{T}(q, p)}{s_\mathcal{S}(q, p)} \right),$$ where $ s(q, p) = \frac{\exp(\cos(q, p) / t)}{\sum_{p' \in P} \exp(\cos(q, p') / t)}$ and $t$ is distillation temperature. In our work, we set $t=1$. 

\section{Offline Experiments}

\subsection{Experiment setup}
To capture broad user intent while adhering strictly to industry norms for data privacy, we relied exclusively on anonymized and aggregated engagement data. Specifically, we sampled 4 million fully anonymized queries with at least one order from a one-year aggregation of system logs. By collecting the top 300 impressed products per query, we safely generated approximately 900 million training pairs.

We evaluated semantic relevance strictly on Exact Match (EM) targets using two distinct index scales: (1) A \textbf{Small Index} of 3.6M products: 122k queries with comprehensive human annotations to measure EM Recall@K. (2) A \textbf{Big Index} of 200M products: 1k traffic-weighted queries to measure EM Precision@K, effectively testing the model's robustness against false positives in a massive search space. We used the small index for rapid iteration and recall-focused analysis, and the big index to better reflect precision behavior under realistic large-scale retrieval conditions. In practice, we often observe strong alignment between EM Recall and EM Precision across models. In this work, we report metrics at $K=20$, noting that evaluations at larger retrieval depths (e.g., $K \in \{128, 256, 512\}$) yield consistent performance trends.

Models were optimized using Adam \citep{adam2014method} with a learning rate of $2e-6$ on 8 NVIDIA A100 GPUs via PyTorch \citep{paszke2019pytorch}. Each query sampled 7 positive products, 3 offline negatives and 5 cross-batch negatives. Distributing a local batch size of 72 across the cluster yielded an effective global batch size of 576. Finally, the model minimized a multi-objective loss (Section \ref{sec:dense}) for $40$ epochs.

\subsection{Offline Experiments Results}
To rigorously quantify the contribution of each proposed enhancement, we conducted a progressive ablation study. By incrementally adding components to the baseline, we isolated their specific impact on retrieval performance. Table \ref{tab:offlineexpr} summarizes these findings as relative improvements over the previous production model. Because all variants incorporate standard in-batch sampling, the reported gains explicitly reflect the additive value of our proposed cross-batch and offline mining strategies. In our production pipeline, offline evaluations serve as directional indicators for rapid iteration. Because these metrics are computed over massive transient pipelines, we rely on large-scale online A/B testing (Section \ref{sec:online_experiment}) to establish strict statistical significance.

\begin{table}[t]
\centering
\caption{Offline results for the progressive ablation study. Performance is reported as relative improvements over the previous production baseline for Data Refresh (\textbf{DR}), Cross-Batch Negatives (\textbf{CB}), Hybrid Hard Negatives (\textbf{HHN}), and Warm-Start Distillation (\textbf{WS}).}\label{tab:offlineexpr}
\setlength{\tabcolsep}{3pt}
\small
\begin{tabular}{lccc}
\toprule
Model 
& \begin{tabular}[c]{@{}c@{}}EM Recall@20\end{tabular}
& \begin{tabular}[c]{@{}c@{}}EM Precision@20\end{tabular} \\
\midrule
DistilBERT proposed in ~\citep{lin2024enhancing} (baseline)  & - & - \\
DistilBERT + DR  & +2.67\%  & +1.60\% \\
DistilBERT + DR + CB  & +2.88\%  & +1.84\% \\
DistilBERT + DR + CB + HHN & +3.15\% & +2.09\% \\
GTE-base + DR + CB + HHN & +3.74\% & +1.97\% \\
GTE-base + DR + CB + HHN + WS & +4.00\% & +2.01\% \\
\bottomrule
\end{tabular}
\end{table}

\subsubsection{Impact of data recency.}
Refreshing the training data (DR) led to consistent baseline improvements, driving relative gains of +2.67\% in EM Recall@20 and +1.60\% in EM Precision@20. These results highlight the critical impact of temporal distribution shift in evolving e-commerce catalogs. However, while data recency establishes a stronger baseline, it does not fundamentally resolve the structural limitations of the dual-encoder, such as its inability to distinguish fine-grained distractors.

\subsubsection{Impact of cross-batch negatives.}
To address the limitations of local batch sizes, we integrated cross-batch (CB) negatives. By sharing embeddings across all distributed GPUs, we exponentially expand the negative pool. This provided an incremental lift over the data-refreshed model, bringing the total EM Recall@20 gain to +2.88\% over the baseline. This suggests that exposing the model to a much larger and more diverse set of negatives significantly improves its discriminative power in the dense vector space.

\subsubsection{Impact of hybrid offline hard negative mining.}
Integrating our hybrid hard negatives (HHN) pushed the cumulative improvements to +3.15\% in EM Recall@20 and +2.09\% in EM Precision@20. Because the baseline already filters basic structural mismatches via PT heuristics, the primary contribution of HHN is its ability to explicitly penalize nuanced semantic distractors. By capturing highly confusable hard negatives that share the correct PT but remain irrelevant, HHN provides strictly verified, high-quality training signals that traditional metadata rules alone cannot surface.

\subsubsection{Impact of encoder capacity and warm-start.}
Finally, we evaluated whether the model architecture itself throttles performance given these enriched training signals. Transitioning from the DistilBERT backbone to the higher-capacity GTE-Base encoder yielded a substantial jump in recall (+3.74\%).
Interestingly, we observed a slight regression in precision (+1.97\% for GTE-base vs. +2.09\% for DistilBERT). This likely reflects a domain-shift penalty: while the larger model’s increased capacity allows it to retrieve more valid targets, initializing it "cold" discards the baseline model's deeply learned boundary constraints. Crucially, applying our Legacy-Aware Warm-Start (WS) initialization successfully recovered this lost precision (+2.01\%) while driving the highest overall recall (+4.00\%). This confirms that transitioning to a high-capacity encoder requires explicit knowledge transfer to fully exploit enriched signals. From a theoretical standpoint, we acknowledge that the success of this Warm-Start Distillation is likely a combined effect. It preserves the deeply learned historical boundary constraints of the legacy system, while simultaneously leveraging the well-documented regularization benefits of knowledge distillation, where the teacher's soft labels provide structural inductive biases that prevent the higher-capacity student from overfitting to false positives.
While the Warm-Start's offline gains (+0.26\% EM Recall@20 and +0.04\% EM Precision@20) appear modest, they represent a massive real-world impact at Walmart's scale. These shifts surface millions of additional relevant products, setting the stage for the live business metrics detailed in Section \ref{sec:online_experiment}.

\section{Online Experiments}
\label{sec:online_experiment}
We deployed our best-performing model (GTE-base with hybrid negatives mining and warm-start) into Walmart's production search environment. We then conducted a two-stage evaluation: a human expert review to assess relevance, followed by large-scale A/B testing to measure business impact. These online results corroborate our offline findings.

\noindent \textbf{Manual Relevance Evaluation.} We performed a side-by-side manual evaluation to ensure the new model met our strict relevance standards. We identified that the new model altered the retrieved top-10 results for 20.85\% of total search traffic compared to the baseline \citep{lin2024enhancing}. We presented the top-10 retrieved items to professional human annotators. As shown in Table~\ref{tab:online_eval}, the proposed model demonstrated a statistically significant lift in ranking quality, achieving a +7.34\% improvement in NDCG@5 and +6.89\% in NDCG@10.

\noindent \textbf{Live A/B Testing.} Following the successful manual evaluation, we conducted A/B testing on live user traffic. The model successfully translated its semantic relevance gains into tangible business value. As shown in Table~\ref{tab:online_eval}, the proposed model delivered a statistically significant gross revenue lift of +0.50\% ($p = 0.03$). 

\begin{table}[ht]
    \centering
    \small 
    \caption{Manual evaluation and online A/B testing results comparing the proposed model against the baseline \citep{lin2024enhancing}.}
    \label{tab:online_eval}
    \begin{tabular}{ccc}
    \toprule
       NDCG@5 Lift (p-val) & NDCG@10 Lift (p-val) & Revenue Lift (p-val)\\
      \midrule
        +7.34\%  (0.00) & +6.89\% (0.00) & +0.50\% (0.03) \\
    \bottomrule
    \end{tabular}
\end{table}


\section{Conclusion}

In this work, we present a unified framework deployed in Walmart's production system that successfully bridges the training-inference gap in large-scale e-commerce retrieval. By scaling negative diversity through an integration of online cross-batch sampling and the novel metadata-aware offline mining, we expose models to a highly refined distribution of challenging distractors. Furthermore, our Legacy-Aware Warm-Start distillation safely mitigates model-evolution risks during capacity upgrades, ensuring that new encoders retain historical domain expertise while fully exploiting enriched training signals. Together, these advancements drive consistent improvements in search relevance and deliver significant tangible business value in live user traffic.

\section*{Presenter Bio}
Hongwei Shang is a Principal Data Scientist at Walmart Global Tech, working on large-scale machine learning systems for e-commerce search, including retrieval and ranking. Her research focuses on embedding-based retrieval, relevance modeling, and knowledge distillation, with an emphasis on bridging offline training and production deployment. She holds a Ph.D. in Statistics from the University of Connecticut.
\bibliographystyle{ACM-Reference-Format}
\bibliography{ANNv20}

@inproceedings{rossi2024relevance,
  title={Relevance filtering for embedding-based retrieval},
  author={Rossi, Nicholas and Lin, Juexin and Liu, Feng and Yang, Zhen and Lee, Tony and Magnani, Alessandro and Liao, Ciya},
  booktitle={Proceedings of the 33rd ACM International Conference on Information and Knowledge Management},
  pages={4828--4835},
  year={2024}
}

@inproceedings{ruan2024idat,
  title={iDAT: inverse distillation adapter-tuning},
  author={Ruan, Jiacheng and Gao, Jingsheng and Xie, Mingye and Dong, Daize and Xiang, Suncheng and Liu, Ting and Fu, Yuzhuo},
  booktitle={2024 IEEE International Conference on Multimedia and Expo (ICME)},
  pages={1--6},
  year={2024},
  organization={IEEE}
}

@inproceedings{devlin2019bert,
  title={Bert: Pre-training of deep bidirectional transformers for language understanding},
  author={Devlin, Jacob and Chang, Ming-Wei and Lee, Kenton and Toutanova, Kristina},
  booktitle={Proceedings of the 2019 conference of the North American chapter of the association for computational linguistics: human language technologies, volume 1 (long and short papers)},
  pages={4171--4186},
  year={2019}
}

@article{johnson2019billion,
  title={Billion-scale similarity search with GPUs},
  author={Johnson, Jeff and Douze, Matthijs and J{\'e}gou, Herv{\'e}},
  journal={IEEE transactions on big data},
  volume={7},
  number={3},
  pages={535--547},
  year={2019},
  publisher={IEEE}
}

@article{kullback1951kullback,
  title={Kullback-leibler divergence},
  author={Kullback, Solomon},
  journal={Encyclopedia of Machine Learning},
  pages={581--583},
  year={1951}
}

@article{mehrdad2024large,
  title={Large language models for relevance judgment in product search},
  author={Mehrdad, Navid and Mohapatra, Hrushikesh and Bagdouri, Mossaab and Chandran, Prijith and Magnani, Alessandro and Cai, Xunfan and Puthenputhussery, Ajit and Yadav, Sachin and Lee, Tony and Zhai, ChengXiang and others},
  journal={arXiv preprint arXiv:2406.00247},
  year={2024}
}

@article{reddy2022shopping,
  title={Shopping queries dataset: A large-scale ESCI benchmark for improving product search},
  author={Reddy, Chandan K and M{\`a}rquez, Llu{\'\i}s and Valero, Fran and Rao, Nikhil and Zaragoza, Hugo and Bandyopadhyay, Sambaran and Biswas, Arnab and Xing, Anlu and Subbian, Karthik},
  journal={arXiv preprint arXiv:2206.06588},
  year={2022}
}

@article{Jiang2023Mistral7,
  title={Mistral 7B},
  author={Albert Qiaochu Jiang and Alexandre Sablayrolles and Arthur Mensch and Chris Bamford and Devendra Singh Chaplot and Diego de Las Casas and Florian Bressand and Gianna Lengyel and Guillaume Lample and Lucile Saulnier and L{\'e}lio Renard Lavaud and Marie-Anne Lachaux and Pierre Stock and Teven Le Scao and Thibaut Lavril and Thomas Wang and Timoth{\'e}e Lacroix and William El Sayed},
  journal={arXiv preprint abs/2310.06825},
  year={2023},
  url={https://arxiv.org/abs/2310.06825}
}

@inproceedings{shang2025knowledge,
  title={Knowledge distillation for enhancing walmart e-commerce search relevance using large language models},
  author={Shang, Hongwei and Vo, Nguyen and Yadav, Nitin and Zhang, Tian and Puthenputhussery, Ajit and Cai, Xunfan and Chen, Shuyi and Chandran, Prijith and Kang, Changsung},
  booktitle={Companion Proceedings of the ACM on Web Conference 2025},
  pages={449--457},
  year={2025}
}

@article{henderson2017efficient,
  title={Efficient natural language response suggestion for smart reply},
  author={Henderson, Matthew and Al-Rfou, Rami and Strope, Brian and Sung, Yun-hsuan and Hegedus, Laszlo and Kurzweil, Ray and others},
  journal={arXiv preprint arXiv:1705.00652},
  year={2017}
}

@article{hinton2015distilling,
  title={Distilling the knowledge in a neural network},
  author={Hinton, Geoffrey and Vinyals, Oriol and Dean, Jeff},
  journal={arXiv preprint arXiv:1503.02531},
  year={2015}
}

@article{hofstatter2020improving,
  title={Improving efficient neural ranking models with cross-architecture knowledge distillation},
  author={Hofst{\"a}tter, Sebastian and Althammer, Sophia and Schr{\"o}der, Michael and Sertkan, Mete and Hanbury, Allan},
  journal={arXiv preprint arXiv:2010.02666},
  year={2020}
}

@article{sanh2019distilbert,
  title={DistilBERT, a distilled version of BERT: smaller, faster, cheaper and lighter},
  author={Sanh, Victor and Debut, Lysandre and Chaumond, Julien and Wolf, Thomas},
  journal={arXiv preprint arXiv:1910.01108},
  year={2019}
}

@inproceedings{lin2024enhancing,
  title={Enhancing Relevance of Embedding-based Retrieval at Walmart},
  author={Lin, Juexin and Yadav, Sachin and Liu, Feng and Rossi, Nicholas and Suram, Praveen R and Chembolu, Satya and Chandran, Prijith and Mohapatra, Hrushikesh and Lee, Tony and Magnani, Alessandro and others},
  booktitle={Proceedings of the 33rd ACM International Conference on Information and Knowledge Management},
  pages={4694--4701},
  year={2024}
}

@article{xiong2020ance,
  title={Approximate Nearest Neighbor Negative Contrastive Learning for Dense Text Retrieval},
  author={Xiong, Lee and Xiong, Chenyan and Li, Ye and Tang, Kwok-Fung and Liu, Jialu and Bennett, Paul and Ahmed, Junaid and Overwijk, Arnold},
  journal={arXiv preprint arXiv:2007.00808},
  year={2020}
}

@inproceedings{magnani2022semantic,
  title={Semantic retrieval at walmart},
  author={Magnani, Alessandro and Liu, Feng and Chaidaroon, Suthee and Yadav, Sachin and Reddy Suram, Praveen and Puthenputhussery, Ajit and Chen, Sijie and Xie, Min and Kashi, Anirudh and Lee, Tony and others},
  booktitle={Proceedings of the 28th ACM SIGKDD Conference on Knowledge Discovery and Data Mining},
  pages={3495--3503},
  year={2022}
}

@inproceedings{qu2021rocketqa,
  title={RocketQA: An optimized training approach to dense passage retrieval for open-domain question answering},
  author={Qu, Yingqi and Ding, Yuchen and Liu, Jing and Liu, Kai and Ren, Ruiyang and Zhao, Wayne Xin and Dong, Daxiang and Wu, Hua and Wang, Haifeng},
  booktitle={Proceedings of the 2021 conference of the North American chapter of the association for computational linguistics: human language technologies},
  pages={5835--5847},
  year={2021}
}

@article{paszke2019pytorch,
  title={Pytorch: An imperative style, high-performance deep learning library},
  author={Paszke, Adam and Gross, Sam and Massa, Francisco and Lerer, Adam and Bradbury, James and Chanan, Gregory and Killeen, Trevor and Lin, Zeming and Gimelshein, Natalia and Antiga, Luca and others},
  journal={Advances in neural information processing systems},
  volume={32},
  year={2019}
}

@article{adam2014method,
  title={A method for stochastic optimization},
  author={Adam, Kingma DP Ba J and others},
  journal={arXiv preprint arXiv:1412.6980},
  volume={1412},
  number={6},
  year={2014}
}

@inproceedings{qiu2022pre,
  title={Pre-training tasks for user intent detection and embedding retrieval in e-commerce search},
  author={Qiu, Yiming and Zhao, Chenyu and Zhang, Han and Zhuo, Jingwei and Li, Tianhao and Zhang, Xiaowei and Wang, Songlin and Xu, Sulong and Long, Bo and Yang, Wen-Yun},
  booktitle={Proceedings of the 31st ACM International Conference on Information \& Knowledge Management},
  pages={4424--4428},
  year={2022}
}

@inproceedings{he2023que2engage,
  title={Que2engage: Embedding-based retrieval for relevant and engaging products at facebook marketplace},
  author={He, Yunzhong and Tian, Yuxin and Wang, Mengjiao and Chen, Feier and Yu, Licheng and Tang, Maolong and Chen, Congcong and Zhang, Ning and Kuang, Bin and Prakash, Arul},
  booktitle={Companion Proceedings of the ACM Web Conference 2023},
  pages={386--390},
  year={2023}
}

@article{freymuth2025hierarchical,
  title={Hierarchical Multi-field Representations for Two-Stage E-commerce Retrieval},
  author={Freymuth, Niklas and Liu, Dong and Ricatte, Thomas and Mansour, Saab},
  journal={arXiv preprint arXiv:2501.18707},
  year={2025}
}

@inproceedings{li2021embedding,
  title={Embedding-based product retrieval in taobao search},
  author={Li, Sen and Lv, Fuyu and Jin, Taiwei and Lin, Guli and Yang, Keping and Zeng, Xiaoyi and Wu, Xiao-Ming and Ma, Qianli},
  booktitle={Proceedings of the 27th ACM SIGKDD Conference on Knowledge Discovery \& Data Mining},
  pages={3181--3189},
  year={2021}
}

@inproceedings{muhamed2023web,
  title={Web-scale semantic product search with large language models},
  author={Muhamed, Aashiq and Srinivasan, Sriram and Teo, Choon-Hui and Cui, Qingjun and Zeng, Belinda and Chilimbi, Trishul and Vishwanathan, SVN},
  booktitle={Pacific-Asia Conference on Knowledge Discovery and Data Mining},
  pages={73--85},
  year={2023},
  organization={Springer}
}

@article{meghwani2025hard,
  title={Hard Negative Mining for Domain-Specific Retrieval in Enterprise Systems},
  author={Meghwani, Hansa and Agarwal, Amit and Pattnayak, Priyaranjan and Patel, Hitesh Laxmichand and Panda, Srikant},
  journal={arXiv preprint arXiv:2505.18366},
  year={2025}
}

@inproceedings{karpukhin2020dense,
  title={Dense Passage Retrieval for Open-Domain Question Answering.},
  author={Karpukhin, Vladimir and Oguz, Barlas and Min, Sewon and Lewis, Patrick SH and Wu, Ledell and Edunov, Sergey and Chen, Danqi and Yih, Wen-tau},
  booktitle={EMNLP (1)},
  pages={6769--6781},
  year={2020}
}

@inproceedings{robertson1994some,
  title={Some simple effective approximations to the 2-poisson model for probabilistic weighted retrieval},
  author={Robertson, Stephen E and Walker, Steve},
  booktitle={SIGIR’94: Proceedings of the Seventeenth Annual International ACM-SIGIR Conference on Research and Development in Information Retrieval, organised by Dublin City University},
  pages={232--241},
  year={1994},
  organization={Springer}
}

@inproceedings{wang2021cross,
  title={Cross-batch negative sampling for training two-tower recommenders},
  author={Wang, Jinpeng and Zhu, Jieming and He, Xiuqiang},
  booktitle={Proceedings of the 44th international ACM SIGIR conference on research and development in information retrieval},
  pages={1632--1636},
  year={2021}
}

@article{li2023towards,
  title={Towards general text embeddings with multi-stage contrastive learning},
  author={Li, Zehan and Zhang, Xin and Zhang, Yanzhao and Long, Dingkun and Xie, Pengjun and Zhang, Meishan},
  journal={arXiv preprint arXiv:2308.03281},
  year={2023}
}

@inproceedings{puthenputhussery2025large,
  title={Large Scale Deployment of BERT Based Cross Encoder Model for Re-Ranking in Walmart Search Engine},
  author={Puthenputhussery, Ajit and Kang, Changsung and Magnani, Alessandro and Zhang, Tian and Shang, Hongwei and Yadav, Nitin and Chandran, Prijith and Madhani, Bhavin and Fu, Yuan-Tai and Wang, He and others},
  booktitle={Proceedings of the 48th International ACM SIGIR Conference on Research and Development in Information Retrieval},
  pages={4365--4369},
  year={2025}
}

@inproceedings{puthenputhussery2026large,
  title={Bert-Based Cross-Encoder for Large-Scale Engagement Prediction and Re-Ranking in Walmart Search Engine},
  author={Chen, Shuyi and Fu, Philip and Puthenputhussery, Ajit and Kang, Changsung and Sun, Ming and Mu, Cun and Yadav, Sachin Kumar and Shang, Hongwei},
  booktitle={Proceedings of the 49th International ACM SIGIR Conference on Research and Development in Information Retrieval},
  year={2026}
}

\end{document}